\documentclass[superscriptaddress, 
reprint,
 amsmath,amssymb,
 aps, prl
]{revtex4-1}
\usepackage{graphicx}
\usepackage{dcolumn}
\usepackage{bm}

\def\<{\langle}
\def\>{\rangle}

\begin{document}

\title{Propulsion of a two-sphere swimmer}
\author{Daphne Klotsa}
\affiliation{School of Physics and Astronomy, University of Nottingham, UK}
\affiliation{Department of Chemistry, Lensfield Rd., University of Cambridge, Cambridge CB2 1EW, UK}
\affiliation{Department of Applied Physical Sciences, University of North Carolina at Chapel Hill, North Carolina 27599-3290, United States}
\author{Kyle A. Baldwin}
\affiliation{School of Physics and Astronomy, University of Nottingham, UK}
\author{Richard J. A. Hill}
\affiliation{School of Physics and Astronomy, University of Nottingham, UK}
\author{R. M. Bowley}
\affiliation{School of Physics and Astronomy, University of Nottingham, UK}
\author{Michael R. Swift}
\affiliation{School of Physics and Astronomy, University of Nottingham, UK}
\date{\today}

\begin{abstract}
We describe experiments and simulations demonstrating the propulsion of a neutrally-buoyant swimmer that consists of a pair of spheres 
attached by a spring, immersed in a vibrating fluid. 
The vibration of the fluid induces relative motion of the spheres which, for sufficiently large amplitudes, can lead to 
motion of the center of mass of the two spheres.
We find that the swimming speed obtained from both experiment and simulation 
agree and collapse onto a single curve if plotted as a function of the streaming Reynolds number,
suggesting that the propulsion is related to streaming flows.
There appears to be a critical
onset value of the streaming Reynolds number for swimming to occur.   
We observe a change in the streaming flows as the Reynolds number increases, from that generated by two independent oscillating spheres
 to a collective flow pattern around the swimmer as a whole. The mechanism for swimming is traced to a strengthening of a jet of fluid in the wake of the swimmer.
\end{abstract}

\maketitle
 
The mechanism by which self-propulsion through a fluid is achieved
has fascinated scientists of many disciplines, 
and the public alike, for aesthetic, practical and fundamental scientific reasons~\cite{vogel,berg,purcell,dance,lauga-powers}. 
In biology and biomechanics the mechanisms behind the way organisms swim gives 
insight into their biological function and purpose~\cite{vogel,berg,goldstein2011,leptos}. 
Recently, the design of efficient ``robots'' able to navigate themselves through various fluids 
has become an important technological and medical challenge that brings together elements of physics, chemistry, 
biology, engineering and fluid mechanics~\cite{dreyfus,williams,bar-cohen}.
Microscopic artificial swimmers have been proposed for use in targeted drug-delivery, see for example 
~\cite{zhang2009,magnetosperm,feng}. 

Purcell's scallop theorem states that at zero Reynolds number an object cannot
swim using a time-reversible stroke: it will end up going back and forth with no net displacement~\cite{purcell}. 
Many types of
small creatures, for example, insects and aquatic invertebrates, swim at intermediate Reynolds numbers (1-100)~\cite{childress2004}. 
In these cases, time-reversal symmetry is broken by non-linearities in the fluid dynamics rather than by the nature of the stroke.
For such swimmers, an interesting question arises: how does the motion evolve as the Reynolds number is increased from zero?
It has been argued that symmetrical objects with symmetrical strokes such as flapping wings 
have an onset for motion at a critical Reynolds number~\cite{vandenberghe2004,alben2005,vandenberghe2006,lu2006},
whereas asymmetrical objects or strokes have a continuous transition as the Reynolds number is increased ~\cite{lauga2007}.

A central problem when designing a practical artificial swimmer
is how to get energy into the system. 
Methods based on electromagnetic or chemical actuation have been developed~\cite{feng} and currently
there is interest in using acoustic techniques to generate propulsion through the
oscillation of entrapped air bubbles~\cite{scallop,ahmed}.
Vladimirov proposed an alternative mechanism that may lead to swimming
based on a deformable object which is neutrally buoyant, but composed 
of coupled spheres with different sizes and densities~\cite{vlad2013}.
Such an object can generate relative motion of its parts if immersed in a vibrating fluid; this
motion may lead to swimming. 
However, his calculations in the absence of fluid and particle inertia predicted that such an object will not 
swim if subjected to unidirectional oscillation.
Here we pose the question: can an experimental realisation of this object be made 
to swim at higher Reynolds numbers, and, if so, what is the 
method of propulsion and the nature of the transition to swimming? 

In this Letter we describe experiments and simulations demonstrating the propulsion of a pair of spheres 
attached by a spring, immersed in a vertically vibrating fluid. 
We consider two particular realisations of this object: one with unequal-sized spheres and the other with equal-sized spheres. 
In both cases, the density of the spheres is different from one another and from the liquid in which they are immersed,
however, the object as a whole is neutrally-buoyant. 
We find that both designs swim for sufficiently high amplitudes of vibration;
the unequal-sized spheres swim upwards, in the direction of the larger, less dense sphere, whereas
the equal-sized spheres swim downwards, in the direction of the higher density sphere.
The data for the swimming speed are found to collapse both in experiment and simulation when scaled appropriately 
with the streaming Reynolds number, suggesting that the streaming flows induced by fluid non-linearities play a 
central role~\cite{riley}. 
Furthermore, the apparent onset of motion appears to be governed by a critical value of the streaming Reynolds number.
The mechanism for propulsion is traced to a change in 
the topology of the streaming flows that transition from those of two noninteracting spheres when the dimer is stationary,
to a collective flow around the object, at the apparent onset of swimming.
The flow field shows a strengthening of a jet of fluid behind the swimmer.

The dimers were constructed from two spheres joined together by a small coil of wire. Examples of the
asymmetric and symmetric dimers are shown in the insets to Figs. 1 and 2 respectively. Details of their
construction and the experimental set-up are given in Supplemental Material~\cite{SM}. The dimers were
designed so that they could be made neutrally buoyant in a salt-water solution. The solution was
vibrated vertically at a given frequency, $f$, amplitude, $A$. 
The dimensionless acceleration of the cell $\Gamma=A(2\pi f)^2/g$ was varied between $1-20$, 
where $g$ is the gravitational acceleration. The
frequency ranged from 30Hz to 135Hz.

As the cell vibrated, each sphere had a different amplitude and phase relative to the fluid motion due to differences 
in the size and/or densities of the two spheres.
At low amplitudes of vibration of the cell the spheres oscillated vertically, but no net time-averaged motion
of the center of mass of the spheres could be observed within experimental error. 
Beyond a certain threshold the dimer started to move; 
increasing the amplitude made the dimer swim faster.

\begin{figure}
\includegraphics[width=\columnwidth]{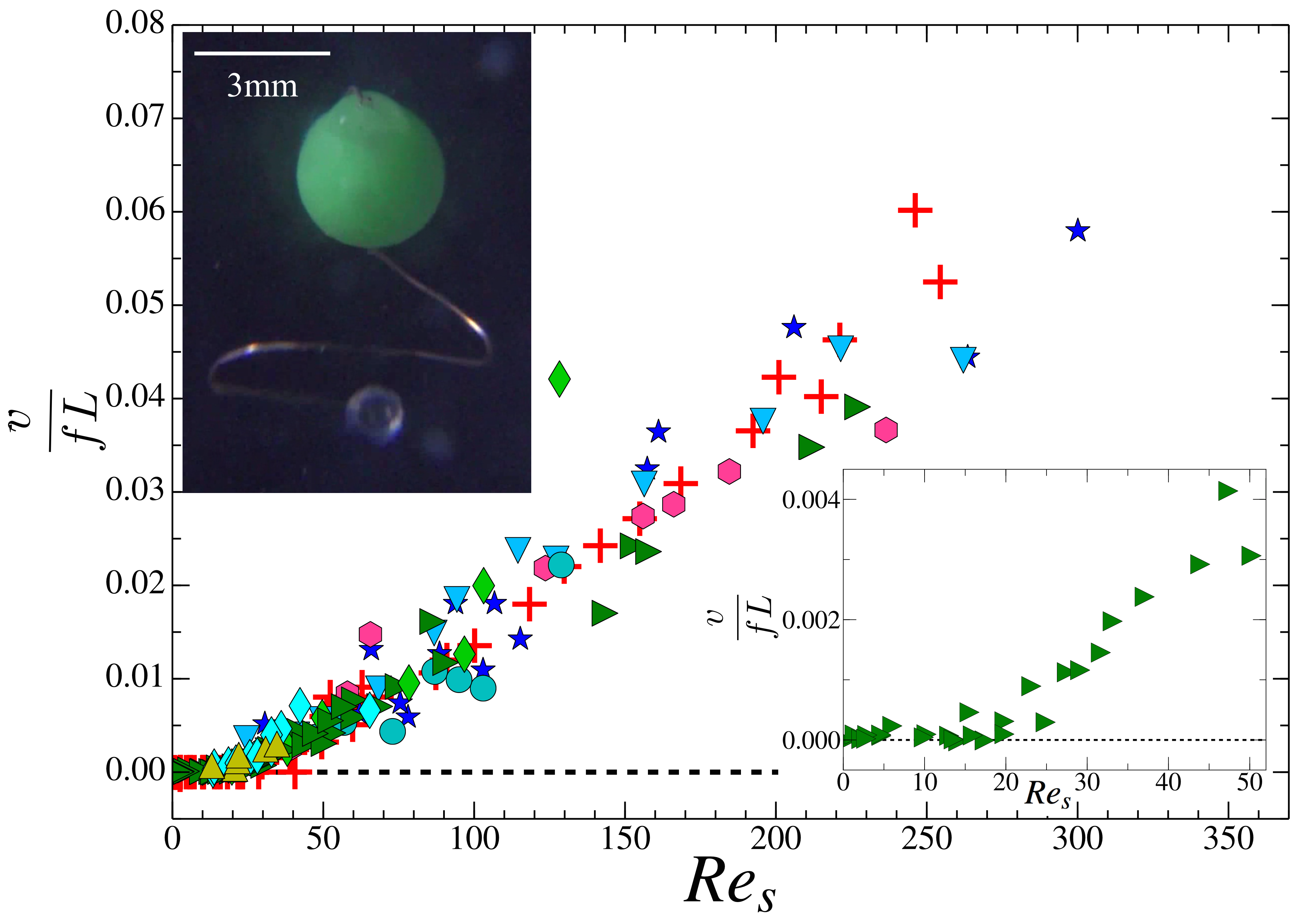}
\caption{Main panel shows the experimental data collapse for the unequal-sized swimmer which swims upwards. The driving conditions
are the following: blue stars $f=65$Hz, light blue triangle down $f=75$Hz, 
pink hexagons $f=85$Hz, 
cyan circles $f=95$Hz, yellow triangle up $f=105$Hz, green triangles right $f=125$Hz, green diamond $f=135$Hz.
In all cases the viscosity was $1.2$mm$^2$/s except for one data set (turquoise diamond $f=135$Hz) where the viscosity was $2.5$mm$^2$/s.
Simulations for $\Gamma$ between $2-20$ and frequencies $f=65, 75, 100, 125$Hz are shown by the red plus symbols for comparison.  
The lower inset shows a close-up of the experimental data ($f=125$Hz) shown in the main panel near the apparent onset.
The upper inset shows a photograph of the swimmer when stationary. 
}
\label{fig:exp-fish}
\end{figure}

To obtain the velocity of the dimer,  
the vibration was initiated abruptly under fixed $\Gamma$ and $f$, and the motion of the dimer was filmed using 
a high-speed camera. 
The movies show that the separation of the spheres varied sinusoidally (indicating that
the coiled wire acted as a linear spring to a good approximation).
From such movies the steady-state velocity of the dimer, $v$, and the relative amplitude of the 
two spheres with respect to each other, $A_r$, was obtained. 
$A_r$ is the amplitude of the relative motion of the two spheres
that comprise the swimmer.
Note that $A_r$ and the driving amplitude of the cell are different;
$A_r$ increases approximately linearly with $A$.
As far as the motion of the spheres is concerned, in the rest frame of the cell, 
$A_r$ and $f$ are the only relevant driving parameters.
As can be seen from the movies~\cite{SM}, the motion of the spheres was predominantly
along the vertical line through their centers; there was very little sideways `waggling' movement.

Fig.~1 shows the data obtained for the two asymmetric dimers, which swim upwards in the direction of the larger
sphere. 
The data collapse (within the scatter) 
when plotted in terms of the dimensionless combinations $v/fL$, and the streaming Reynolds number $Re_s=A_r^2/\delta^2$. Here $L$ is the
diameter of the larger sphere and $\delta=(\nu/2\pi f)^\frac{1}{2}$ is the viscous length in terms of the kinematic viscosity $\nu$.
$Re_s$ is one of three dimensionless ratios that can be defined from the length scales $A_r$, $L$ and $\delta$
and characterises the time-averaged (steady) flow~\cite{riley,otto}. In our experiments $L\gg\delta$ which results in
a configuration of the time-averaged flow around each sphere that has inner and outer loops~\cite{riley,klotsa1}.
The data are consistent between the measurements obtained from two nominally identical, asymmetric dimers, 
indicating that small differences in construction such as variations in the shape of the loop of wire and of 
the shape and amount of glue have little effect.
The collapse in terms of $Re_s$ shows that the motion is related to streaming flows generated by the vibration
of the dimer.
The lower inset shows data taken at low amplitudes of vibration
and suggests a sharp increase in velocity at $Re_s \approx 20$.

Fig.~2 shows the behaviour of the equal-sized dimer. 
In this case it moves in
the opposite direction, downwards, with the heavier sphere at the front. 
The data illustrates that the speed and direction of motion 
depends on the densities and sizes of the spheres, as well as the gap
between them; if the two spheres are sufficiently far apart, the dimer will not swim.
The main reason for considering
this system is that it is arguably one of the simplest objects that can be made to swim.
Note that it was difficult to design dimers made of equal-sized spheres of different
densities that can be made neutrally buoyant in salt solutions, and have
sufficient mass difference between the spheres to generate enough relative motion to induce swimming. 
Hence the relative paucity of data compared to that obtained for the unequal-sized dimers.

\begin{figure}
\includegraphics[width=\columnwidth]{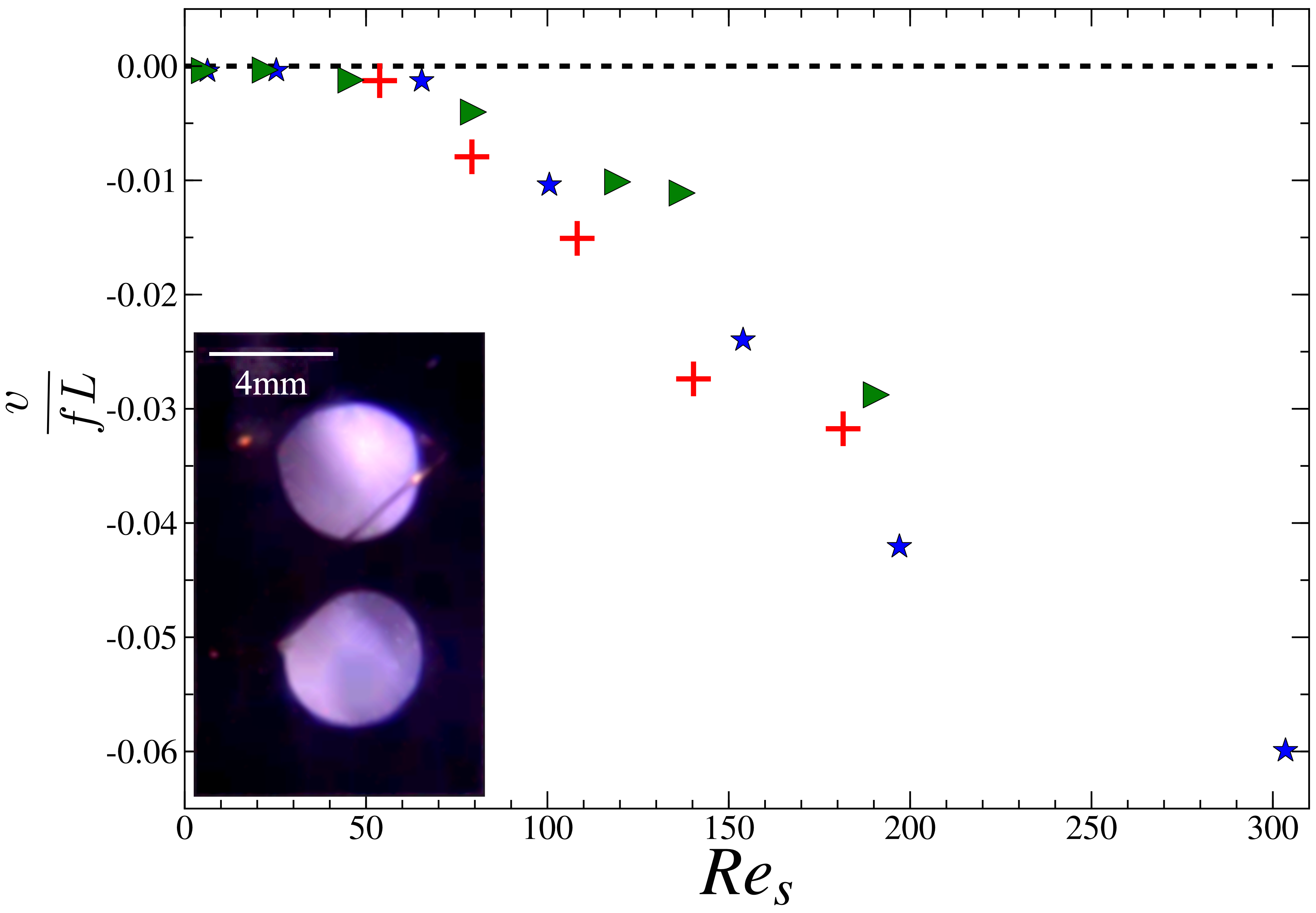}
\caption{Main panel shows the experimental data collapse for the equal-sized swimmer. 
The driving conditions are the following: blue stars $f=30$Hz and green triangles right $f=35$Hz. 
Simulations for $\Gamma$ between $2-12$ and frequency $f=30$Hz are shown by the red plus symbols for comparison.
The inset shows a photograph of the swimmer when stationary.}
\label{fig:equal}
\end{figure}

\begin{figure}[h!]
\includegraphics[width=70mm]{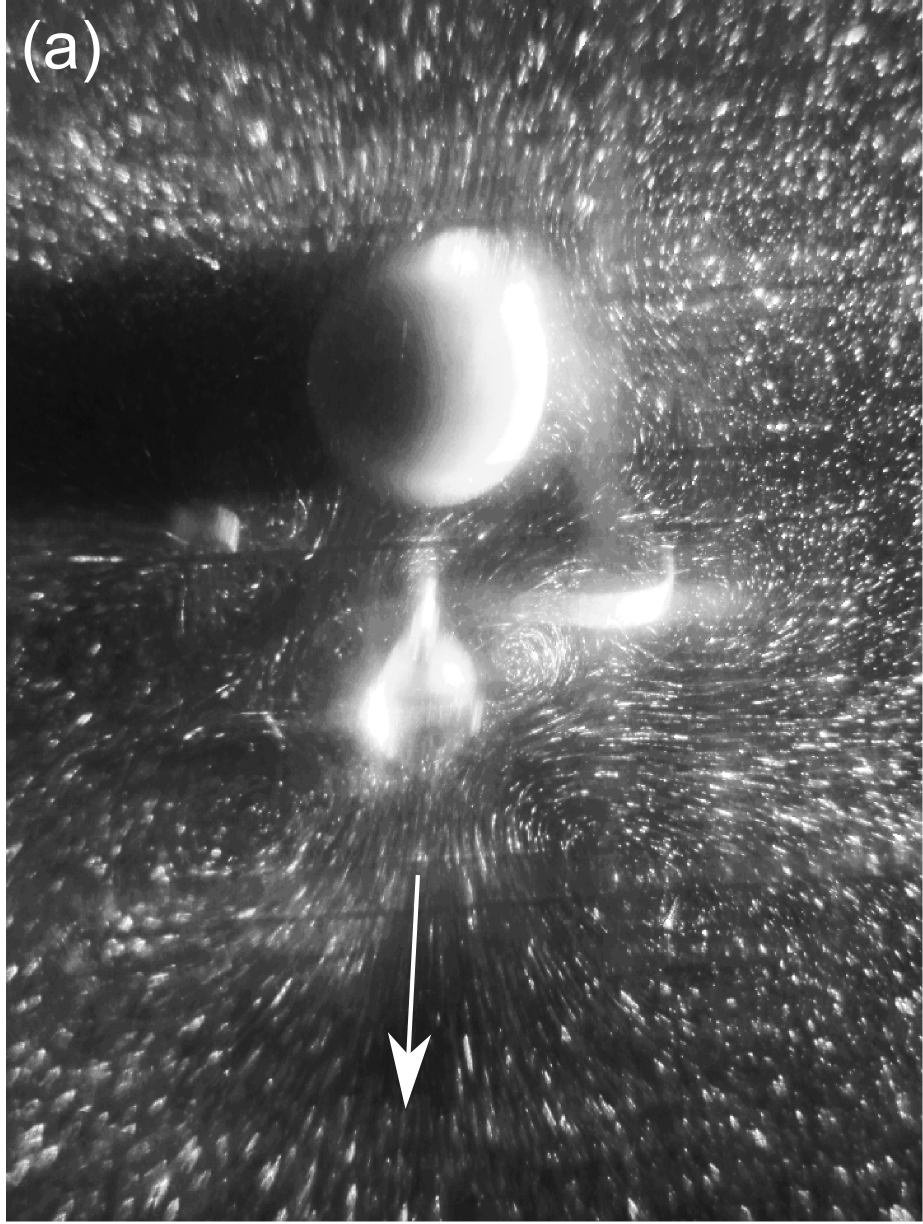}
\includegraphics[width=70mm]{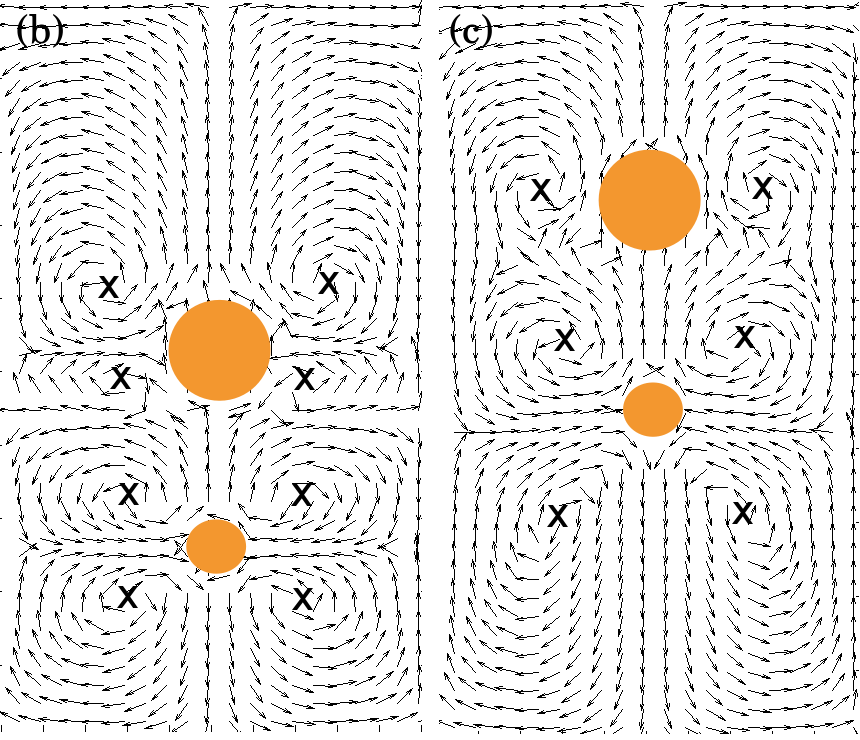}
\caption{Illustrations of the fluid flows generated by the vibration of the spheres from experiment and simulation. 
Panel (a) shows an image taken from experiment showing the flow around the spheres. The arrow illustrates the direction of
a jet of fluid evident from the movies (supplemental information~\cite{SM}).
Panels (b) and (c) show the direction of
the time-averaged velocity field (i.e. the \emph{normalised} velocity vectors) in the plane of the spheres. In (b) the swimmer is stationary ($Re_s=15$)
while in panel (c) it is swimming
($Re_s=60$). These figures illustrate the change in topology of the flows as the amplitude
of vibration increases. Note that the magnitude of the flows is much greater around the
smaller sphere than around the larger sphere, as seen in panel (a).}
\label{fig:flows}
\end{figure}

\begin{figure}
\includegraphics[width=\columnwidth]{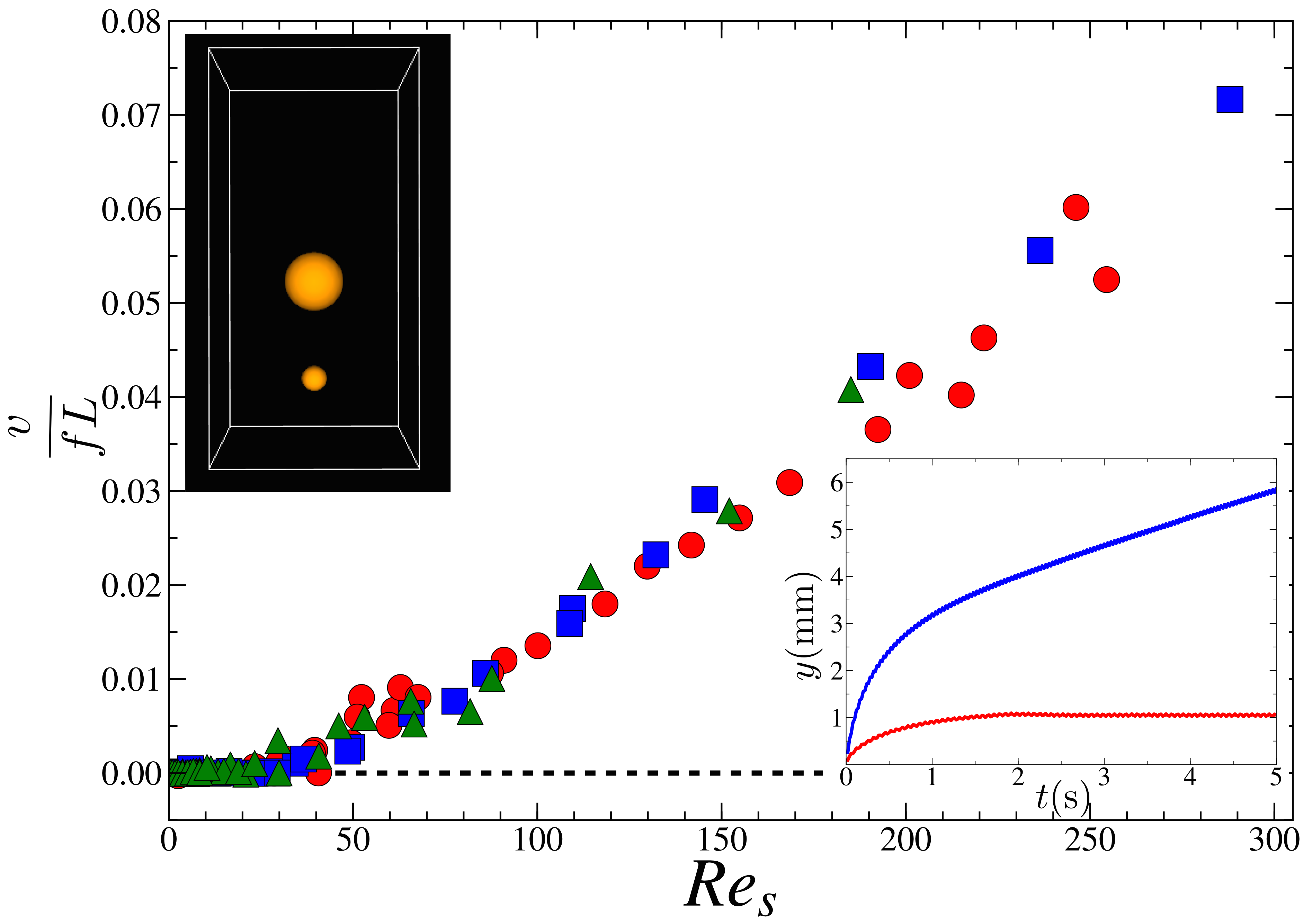}
\caption{Data collapse from simulations confirming the scaling behaviour for different viscosities (red 
$1.2\times 10^{-6} $m$^2$ /s, blue $2\times 10^{-6} $m$^2$/s, green $3\times 10^{-6} $m$^2$/s). Each data set includes simulations for
$\Gamma$ between $2-20$ and $f=65, 75, 100,125$Hz. 
The lower inset shows typical trajectories of a swimmer that stays 
stationary for $\Gamma=2$, $f=65$Hz (red line)
and one that swims for $\Gamma=12$, $f=75$Hz (blue line).
The upper left inset is a snapshot from simulations showing the swimmer
and the simulated cell.}
\label{fig:sim}
\end{figure}

We now ask the question: what is the cause of the motion?  To address this we first imaged the flow using tracer particles illuminated by a planar laser sheet in the plane containing the centres of the two spheres of the dimer. Fig.~3~(a) shows a photograph of 
the asymmetric dimer taken with an exposure time of one period of oscillation, revealing the motion of the tracer particles. In this image
the dimer is close to the onset of motion.  A 
downward jet originates from the vicinity of the lower sphere~\cite{SM}.
Similar behaviour was found for equal-sized spheres, except that the strong jet was generated by the upper, 
lighter sphere, causing the swimmer to swim downwards.  

In order to investigate the motion of the spheres and the fluid in detail we used simulations 
which were based on an embedded boundary method described previously~\cite{klotsa1,wright,klotsa2,klotsa3,
hector}.
The fluid was assumed to obey the Navier-Stokes equations which were discretised on a staggered mesh~\cite{mac}
and solved using the projection method~\cite{projection} to ensure incompressibility of the fluid.
The interaction between the fluid and the rigid spheres
was achieved through the template model, which introduces a two-way coupling between the particles and the fluid~\cite{klotsa3}. 
The spheres were joined by a linear spring as in the experiments.
An equal and opposite force was applied vertically to the spheres to mimic the effects of static buoyancy,
rather than imposing the effect of gravity directly on the fluid. 
The influence of vibration was introduced by applying a sinusoidal acceleration to the fluid and particles, 
so that the simulations were carried out in the frame of reference of the vibrated cell.

The computational parameters of the swimmer (size, density and gap) and fluid (viscosity and density) were chosen 
to match the experiments. However, 
any interaction of the wire with the fluid was ignored and the dimers were assumed to be made of perfect spheres.
Details of the parameters used are given in the Supplemental Material.
One difference between experiment and simulation is that the simulated cells are smaller due to computational limitations.
Examples of the simulated data are shown in Figs. 1 and 2 by the large red plus symbols. 
There is clearly good agreement between
the simulations and experiment despite the numerical limitations arising from the simulated cell size
and possible fluid lattice effects.

The simulations allow us to determine in more detail the fluid flows generated by the motion of the spheres induced by the vibration of
the cell. This flow is best illustrated by plotting the direction of the velocity field in the vertical plane through
the center of the two spheres. Examples of these flows for the two unequal-sized spheres, time-averaged over a cycle,
are shown in Fig.~3~(b) and (c). At low amplitudes, Fig.~3~(b), there are two outer vortex rings around each sphere,
marked by crosses. 
This is the flow pattern expected if the flows of the two spheres 
do not interact strongly~\cite{otto}. 
Under these conditions the time-averaged center of mass of the two spheres remains stationary:
the dimer does not swim.

As the amplitude increases, the flows grow in strength, but more importantly, the flows around each sphere
start to interact strongly. 
The lower loop of the upper sphere is forced towards the surface of the sphere and reduces in size.
Eventually, for sufficiently high amplitudes, there are only three vortex loops, as shown in Fig.~3~(c).
A jet of fluid directed downwards
from the smaller sphere can be observed from the plot of the normalised velocity field, Fig.~3~(c), and from experiment Fig.~3~(a). 
Under these conditions, the swimmer moves upwards, in the opposite direction to the jet.

Simulations also allow us to vary parameters which are not easily accessible experimentally, such as 
a wider range of fluid viscosities, as shown in Fig.~4. When the dimer is moving there are four independent length scales:
$v/f$, $A_r$, $L$ and the viscous length $\delta$. We obtain the best
data collapse if $v/f$ is made non-dimensional by dividing by $L$ rather than either of 
the other two length scales (see Supplemental Material \cite{SM}).
Fig.~4 shows the simulation data plotted in this way indicating data collapse,
the same way as  the experimental data collapse shown in Fig. 1.
The lower right inset to Fig. 4. shows typical trajectories after vibration has been applied. There are a few seconds of
transient motion before the steady-state velocity is reached.

Figs. 1, 2 and 4 all show that $v/fL$ scales approximately linearly with the streaming Reynolds number $Re_s$ for sufficiently large amplitudes $A_r$. This
behaviour is different from that observed for magnetic granular snakes~\cite{belkin} and rigid dimers on surfaces~\cite{wright}.
A simple argument can be constructed to explain the scaling behaviour.
Taking the unequal-sized swimmer as an example, the smaller sphere has a much larger amplitude of motion than the larger sphere, (see movie in the Supplemental Material~\cite{SM}).
The smaller sphere acts as a pump, imparting downward momentum to the fluid. 
The reaction force on the small sphere is equal and opposite to the rate of momentum transfer to the fluid. Its magnitude 
is proportional to the square of its speed $(fA_r)^2$ , the fluid density, $\rho$, and a geometric factor proportional to $L^2$.
In this simple model, the force is balanced by the Stokes' drag on the larger sphere which scales as $6\pi L\eta v$ with $v$ the velocity of the swimmer
and $\eta$ is the dynamic viscosity of the fluid ($\rho \nu$). 
By equating the two forces we obtain $v/fL$ proportional to $Re_s=A_r^2/\delta^2$ as observed in the data for large amplitudes. 

Note, however, that this particular scaling behaviour is not expected to hold
generally as there are four independent length scales in this problem, and therefore three independent
dimensionless ratios of lengths. The argument presented above is only expected to hold in the limit $L \gg \delta$.

The analysis given above assumes a strong asymmetry of the flows around both spheres,
an assumption that breaks down at lower Reynolds numbers, as shown from the flow patterns in Fig.~3.
In both experiment and simulation there {\em appears} to be a critical onset
value of $Re_s \approx 20$ for swimming to occur, obtained by extrapolation of the
data to $v=0$. 
It has been argued that asymmetric objects have a continuous transition to swimming \cite{lauga2007}.
This is not necessarily inconsistent with our observations.
For $Re_s$ below the apparent onset it is difficult to determine whether $v$ is strictly zero or is just small:
experimentally it is hard to ensure that any small centre-of-mass motion is not
due to residual buoyancy; 
in simulation, lattice effects may influence the motion when the amplitudes of the spheres
become comparable to the lattice spacing.
The existence of an apparent onset to motion has also been observed in an 
asymmetric flapping wing~\cite{vandenberghe2006}
and the `acoustic scallop'~\cite{scallop}. 
The good agreement between the experiment and simulation for our system allows us to conjecture that
the apparent onset of motion arises from the change in topology of the streaming flows.

The examples presented here show a rich variety of behaviour but only represent a small part of the parameter space. A systematic
investigation into the influence of the overall size of the dimer, the ratio of the sphere diameters, the sphere density ratios and the gap width would be informative.
It would be of interest to make a fully self-propelled swimmer based on the relative
vibration of two spheres, driven by an internal linear motor, since such swimmers would not be constrained to move along one axis.
Collections of such swimmers might be expected to exhibit interesting cooperative behaviour induced by interacting streaming flows~\cite{wun,voth,klotsa1,klotsa2,klotsa3}.

\acknowledgements
D.K. would like to thank Sharon Glotzer for support and guidance.
D.K. acknowledges FP7 Marie Curie Actions of the European Commission (PIOF-GA-2011-302490 Actsa). 
R.J.A.H. acknowledges support from an EPSRC Fellowship; Grant No. EP/I004599/1.


\begin{thebibliography}{10}%
\makeatletter
\providecommand \@ifxundefined [1]{%
 \ifx #1\undefined \expandafter \@firstoftwo
 \else \expandafter \@secondoftwo
\fi
}%
\providecommand \@ifnum [1]{%
 \ifnum #1\expandafter \@firstoftwo
 \else \expandafter \@secondoftwo
\fi
}%
\providecommand \enquote [1]{``#1''}%
\providecommand \bibnamefont  [1]{#1}%
\providecommand \bibfnamefont [1]{#1}%
\providecommand \citenamefont [1]{#1}%
\providecommand\href[0]{\@sanitize\@href}%
\providecommand\@href[1]{\endgroup\@@startlink{#1}\endgroup\@@href}%
\providecommand\@@href[1]{#1\@@endlink}%
\providecommand \@sanitize [0]{\begingroup\catcode`\&12\catcode`\#12\relax}%
\@ifxundefined \pdfoutput {\@firstoftwo}{%
 \@ifnum{\z@=\pdfoutput}{\@firstoftwo}{\@secondoftwo}%
}{%
 \providecommand\@@startlink[1]{\leavevmode}%
 \providecommand\@@endlink[0]{}%
}{%
 \providecommand\@@startlink[1]{%
  \leavevmode
  \pdfstartlink
   attr{/Border[0 0 1 ]/H/I/C[0 1 1]}%
   user{/Subtype/Link/A<</Type/Action/S/URI/URI(#1)>>}%
  \relax
 }%
 \providecommand\@@endlink[0]{\pdfendlink}%
}%
\providecommand \url  [0]{\begingroup\@sanitize \@url }%
\providecommand \@url [1]{\endgroup\@href {#1}{\urlprefix}}%
\providecommand \urlprefix [0]{URL }%
\providecommand \Eprint[0]{\href }%
\@ifxundefined \urlstyle {%
  \providecommand \doi [1]{doi:\discretionary{}{}{}#1}%
}{%
  \providecommand \doi [0]{doi:\discretionary{}{}{}\begingroup
  \urlstyle{rm}\Url }%
}%
\providecommand \doibase [0]{http://dx.doi.org/}%
\providecommand \Doi[1]{\href{\doibase#1}}%
\providecommand \bibAnnote [3]{%
  \BibitemShut{#1}%
  \begin{quotation}\noindent
    \textsc{Key:}\ #2\\\textsc{Annotation:}\ #3%
  \end{quotation}%
}%
\providecommand \bibAnnoteFile [2]{%
  \IfFileExists{#2}{\bibAnnote {#1} {#2} {\input{#2}}}{}%
}%
\providecommand \typeout [0]{\immediate \write \m@ne }%
\providecommand \selectlanguage [0]{\@gobble}%
\providecommand \bibinfo [0]{\@secondoftwo}%
\providecommand \bibfield [0]{\@secondoftwo}%
\providecommand \translation [1]{[#1]}%
\providecommand \BibitemOpen[0]{}%
\providecommand \bibitemStop [0]{}%
\providecommand \bibitemNoStop [0]{.\EOS\space}%
\providecommand \EOS [0]{\spacefactor3000\relax}%
\providecommand \BibitemShut [1]{\csname bibitem#1\endcsname}%
\bibitem{vogel}%
  \BibitemOpen
  \bibfield{author}{%
  \bibinfo {author} {\bibfnamefont{S.}~\bibnamefont{Vogel}},\ }%
  \emph{\bibinfo {title} {Life in Moving Fluids: The Physical Biology of
  Flow}}\ (\bibinfo {publisher} {Princeton University Press},\ \bibinfo {year}
  {1996})%
  \bibAnnoteFile{NoStop}{vogel}%
\bibitem{berg}%
  \BibitemOpen
  \bibfield{author}{%
  \bibinfo {author} {\bibfnamefont{H.}~\bibnamefont{Berg}},\ }%
  \bibfield{journal}{%
  \bibinfo {journal} {Physics Today}\ }%
  \textbf{\bibinfo {volume} {53}},\ \bibinfo {pages} {24} (\bibinfo {year}
  {2000})%
  \bibAnnoteFile{NoStop}{berg}%
\bibitem{purcell}%
  \BibitemOpen
  \bibfield{author}{%
  \bibinfo {author} {\bibfnamefont{E.~M.}\ \bibnamefont{Purcell}},\ }%
  \bibfield{journal}{%
  \bibinfo {journal} {Am. J. Phys.}\ }%
  \textbf{\bibinfo {volume} {45}},\ \bibinfo {pages} {3} (\bibinfo {year}
  {1977})%
  \bibAnnoteFile{NoStop}{purcell}%
\bibitem{dance}%
  \BibitemOpen
  \bibfield{author}{%
  \bibinfo {author} {\bibfnamefont{E.}~\bibnamefont{Lauga}}\ and\ \bibinfo
  {author} {\bibfnamefont{R.~E.}\ \bibnamefont{Goldstein}},\ }%
  \bibfield{journal}{%
  \bibinfo {journal} {Physics Today}\ }%
  \textbf{\bibinfo {volume} {65}},\ \bibinfo {pages} {30} (\bibinfo {year}
  {2012})%
  \bibAnnoteFile{NoStop}{dance}%
\bibitem{lauga-powers}%
  \BibitemOpen
  \bibfield{author}{%
  \bibinfo {author} {\bibfnamefont{E.}~\bibnamefont{Lauga}}\ and\ \bibinfo
  {author} {\bibfnamefont{T.~R.}\ \bibnamefont{Powers}},\ }%
  \bibfield{journal}{%
  \bibinfo {journal} {Rep. Prog. Phys.}\ }%
  \textbf{\bibinfo {volume} {72}},\ \bibinfo {pages} {096601} (\bibinfo {year}
  {2009})%
  \bibAnnoteFile{NoStop}{lauga-powers}%
\bibitem{goldstein2011}%
  \BibitemOpen
  \bibfield{author}{%
  \bibinfo {author} {\bibfnamefont{R.~E.}\ \bibnamefont{Goldstein}}, \bibinfo
  {author} {\bibfnamefont{M.}~\bibnamefont{Polin}},\ and\ \bibinfo {author}
  {\bibfnamefont{I.}~\bibnamefont{Tuval}},\ }%
  \bibfield{journal}{%
  \bibinfo {journal} {Phys. Rev. Lett.}\ }%
  \textbf{\bibinfo {volume} {107}},\ \bibinfo {pages} {148103} (\bibinfo {year}
  {2011})%
  \bibAnnoteFile{NoStop}{goldstein2011}%
\bibitem{leptos}%
  \BibitemOpen
  \bibfield{author}{%
  \bibinfo {author} {\bibfnamefont{K.~C.}\ \bibnamefont{Leptos}}, \bibinfo
  {author} {\bibfnamefont{K.~Y.}\ \bibnamefont{Wan}}, \bibinfo {author}
  {\bibfnamefont{M.}~\bibnamefont{Polin}}, \bibinfo {author}
  {\bibfnamefont{I.}~\bibnamefont{Tuval}}, \bibinfo {author}
  {\bibfnamefont{A.~I.}\ \bibnamefont{Pesci}},\ and\ \bibinfo {author}
  {\bibfnamefont{R.~E.}\ \bibnamefont{Goldstein}},\ }%
  \bibfield{journal}{%
  \bibinfo {journal} {Phys. Rev. Lett.}\ }%
  \textbf{\bibinfo {volume} {111}},\ \bibinfo {pages} {158101} (\bibinfo {year}
  {2013})%
  \bibAnnoteFile{NoStop}{leptos}%
\bibitem{dreyfus}%
  \BibitemOpen
  \bibfield{author}{%
  \bibinfo {author} {\bibfnamefont{R.}~\bibnamefont{Dreyfus}}, \bibinfo
  {author} {\bibfnamefont{J.}~\bibnamefont{Baudry}}, \bibinfo {author}
  {\bibfnamefont{M.~L.}\ \bibnamefont{Roper}}, \bibinfo {author}
  {\bibfnamefont{M.}~\bibnamefont{Fermigier}}, \bibinfo {author}
  {\bibfnamefont{H.~A.}\ \bibnamefont{Stone}},\ and\ \bibinfo {author}
  {\bibfnamefont{J.}~\bibnamefont{Bibette}},\ }%
  \bibfield{journal}{%
  \bibinfo {journal} {Nature}\ }%
  \textbf{\bibinfo {volume} {437}},\ \bibinfo {pages} {862} (\bibinfo {year}
  {2005})%
  \bibAnnoteFile{NoStop}{dreyfus}%
\bibitem{williams}%
  \BibitemOpen
  \bibfield{author}{%
  \bibinfo {author} {\bibfnamefont{B.~J.}\ \bibnamefont{Williams}}, \bibinfo
  {author} {\bibfnamefont{S.~V.}\ \bibnamefont{Anand}}, \bibinfo {author}
  {\bibfnamefont{J.}~\bibnamefont{Rajagopalan}},\ and\ \bibinfo {author}
  {\bibfnamefont{M.~T.~A.}\ \bibnamefont{Saif}},\ }%
  \bibfield{journal}{%
  \bibinfo {journal} {Nature Comm.}\ }%
  \textbf{\bibinfo {volume} {5}},\ \bibinfo {pages} {3081} (\bibinfo {year}
  {2013})%
  \bibAnnoteFile{NoStop}{williams}%
\bibitem{bar-cohen}%
  \BibitemOpen
  \bibfield{author}{%
  \bibinfo {author} {\bibfnamefont{Y.}~\bibnamefont{Bar-Cohen}},\ }%
  \bibfield{journal}{%
  \bibinfo {journal} {Bioinspir. Biomim.}\ }%
  \textbf{\bibinfo {volume} {1}},\ \bibinfo {pages} {P1} (\bibinfo {year}
  {2006})%
  \bibAnnoteFile{NoStop}{bar-cohen}%
\bibitem{zhang2009}%
  \BibitemOpen
  \bibfield{author}{%
  \bibinfo {author} {\bibfnamefont{L.}~\bibnamefont{Zhang}}, \bibinfo {author}
  {\bibfnamefont{J.~J.}\ \bibnamefont{Abbott}}, \bibinfo {author}
  {\bibfnamefont{L.}~\bibnamefont{Dong}}, \bibinfo {author}
  {\bibfnamefont{B.~E.}\ \bibnamefont{Kratochvil}}, \bibinfo {author}
  {\bibfnamefont{D.}~\bibnamefont{Bell}},\ and\ \bibinfo {author}
  {\bibfnamefont{B.~J.}\ \bibnamefont{Nelson}},\ }%
  \bibfield{journal}{%
  \bibinfo {journal} {App. Phys. Lett.}\ }%
  \textbf{\bibinfo {volume} {94}},\ \bibinfo {pages} {064107} (\bibinfo {year}
  {2009})%
  \bibAnnoteFile{NoStop}{zhang2009}%
\bibitem{magnetosperm}%
  \BibitemOpen
  \bibfield{author}{%
  \bibinfo {author} {\bibfnamefont{I.~S.~M.}\ \bibnamefont{Khalil}}, \bibinfo
  {author} {\bibfnamefont{H.~C.}\ \bibnamefont{Dijkslag}}, \bibinfo {author}
  {\bibfnamefont{L.}~\bibnamefont{Abelmann}},\ and\ \bibinfo {author}
  {\bibfnamefont{S.}~\bibnamefont{Misra}},\ }%
  \bibfield{journal}{%
  \bibinfo {journal} {App. Phys. Lett.}\ }%
  \textbf{\bibinfo {volume} {104}},\ \bibinfo {pages} {223701} (\bibinfo {year}
  {2014})%
  \bibAnnoteFile{NoStop}{magnetosperm}%
\bibitem{feng}%
  \BibitemOpen
  \bibfield{author}{%
  \bibinfo {author} {\bibfnamefont{J.}~\bibnamefont{Feng}}\ and\ \bibinfo
  {author} {\bibfnamefont{S.~K.}\ \bibnamefont{Cho}},\ }%
  \bibfield{journal}{%
  \bibinfo {journal} {Micromachines}\ }%
  \textbf{\bibinfo {volume} {5}},\ \bibinfo {pages} {97} (\bibinfo {year}
  {2014})%
  \bibAnnoteFile{NoStop}{feng}%
\bibitem{childress2004}%
  \BibitemOpen
  \bibfield{author}{%
  \bibinfo {author} {\bibfnamefont{S.}~\bibnamefont{Childress}}\ and\ \bibinfo
  {author} {\bibfnamefont{R.}~\bibnamefont{Dudley}},\ }%
  \bibfield{journal}{%
  \bibinfo {journal} {J. Fluid Mech.}\ }%
  \textbf{\bibinfo {volume} {498}},\ \bibinfo {pages} {257} (\bibinfo {year}
  {2004})%
  \bibAnnoteFile{NoStop}{childress2004}%
\bibitem{vandenberghe2004}%
  \BibitemOpen
  \bibfield{author}{%
  \bibinfo {author} {\bibfnamefont{N.}~\bibnamefont{Vandenberghe}}, \bibinfo
  {author} {\bibfnamefont{J.}~\bibnamefont{Zhang}},\ and\ \bibinfo {author}
  {\bibfnamefont{S.}~\bibnamefont{Childress}},\ }%
  \bibfield{journal}{%
  \bibinfo {journal} {J. Fluid Mech.}\ }%
  \textbf{\bibinfo {volume} {506}},\ \bibinfo {pages} {147} (\bibinfo {year}
  {2004})%
  \bibAnnoteFile{NoStop}{vandenberghe2004}%
\bibitem{alben2005}%
  \BibitemOpen
  \bibfield{author}{%
  \bibinfo {author} {\bibfnamefont{S.}~\bibnamefont{Alben}}\ and\ \bibinfo
  {author} {\bibfnamefont{M.}~\bibnamefont{Shelley}},\ }%
  \bibfield{journal}{%
  \bibinfo {journal} {PNAS}\ }%
  \textbf{\bibinfo {volume} {102}},\ \bibinfo {pages} {11163} (\bibinfo {year}
  {2005})%
  \bibAnnoteFile{NoStop}{alben2005}%
\bibitem{vandenberghe2006}%
  \BibitemOpen
  \bibfield{author}{%
  \bibinfo {author} {\bibfnamefont{N.}~\bibnamefont{Vandenberghe}}, \bibinfo
  {author} {\bibfnamefont{S.}~\bibnamefont{Childress}},\ and\ \bibinfo {author}
  {\bibfnamefont{J.}~\bibnamefont{Zhang}},\ }%
  \bibfield{journal}{%
  \bibinfo {journal} {Physics of Fluids}\ }%
  \textbf{\bibinfo {volume} {18}},\ \bibinfo {pages} {014102} (\bibinfo {year}
  {2006})%
  \bibAnnoteFile{NoStop}{vandenberghe2006}%
\bibitem{lu2006}%
  \BibitemOpen
  \bibfield{author}{%
  \bibinfo {author} {\bibfnamefont{L.}~\bibnamefont{Xi-Yun}}\ and\ \bibinfo
  {author} {\bibfnamefont{L.}~\bibnamefont{Qin}},\ }%
  \bibfield{journal}{%
  \bibinfo {journal} {Phys. Fluids}\ }%
  \textbf{\bibinfo {volume} {18}},\ \bibinfo {pages} {098104} (\bibinfo {year}
  {2006})%
  \bibAnnoteFile{NoStop}{lu2006}%
\bibitem{lauga2007}%
  \BibitemOpen
  \bibfield{author}{%
  \bibinfo {author} {\bibfnamefont{E.}~\bibnamefont{Lauga}},\ }%
  \bibfield{journal}{%
  \bibinfo {journal} {Phys. Fluids}\ }%
  \textbf{\bibinfo {volume} {19}},\ \bibinfo {pages} {061703} (\bibinfo {year}
  {2007})%
  \bibAnnoteFile{NoStop}{lauga2007}%
\bibitem{scallop}%
  \BibitemOpen
  \bibfield{author}{%
  \bibinfo {author} {\bibfnamefont{R.~J.}\ \bibnamefont{Dijkinc}}, \bibinfo
  {author} {\bibfnamefont{J.~P.}\ \bibnamefont{van der Dennen}}, \bibinfo {author}
  {\bibfnamefont{C.~D.}~\bibnamefont{Ohl}},\ and\ \bibinfo {author}
  {\bibfnamefont{A.}~\bibnamefont{Prosperetti}},\ }%
  \bibfield{journal}{%
  \bibinfo {journal} {J. Micromech. Microeng.}\ }%
  \textbf{\bibinfo {volume} {16}},\ \bibinfo {pages} {1653} (\bibinfo {year}
  {2006})%
  \bibAnnoteFile{NoStop}{scallop}%
\bibitem{ahmed}%
  \BibitemOpen
  \bibfield{author}{%
  \bibinfo {author} {\bibfnamefont{D.}\ \bibnamefont{Ahmed}}, \bibinfo
  {author} {\bibfnamefont{M.}\ \bibnamefont{Lu}}, \bibinfo {author}
  {\bibfnamefont{A.}~\bibnamefont{Nourhani}}, \bibinfo {author}
  {\bibfnamefont{P.~E.}~\bibnamefont{Lammert}}, \bibinfo {author}
  {\bibfnamefont{Z.}\ \bibnamefont{Stratton}},
\bibinfo {author}
  {\bibfnamefont{H.~S.}\ \bibnamefont{Muddana}},\bibinfo {author}
  {\bibfnamefont{V.~H.}\ \bibnamefont{Crespi}},
\ and\ \bibinfo {author}
  {\bibfnamefont{T.~J.}\ \bibnamefont{Huang}},\ }%
  \bibfield{journal}{%
  \bibinfo {journal} {Scientific Reports}\ }%
  \textbf{\bibinfo {volume} {5}},\ \bibinfo {pages} {9744} (\bibinfo {year}
  {2015})%
  \bibAnnoteFile{NoStop}{ahmed}%
\bibitem{vlad2013}%
  \BibitemOpen
  \bibfield{author}{%
  \bibinfo {author} {\bibfnamefont{V.~A.}\ \bibnamefont{Vladimirov}},\ }%
  \bibfield{journal}{%
  \bibinfo {journal} {J. Fluid Mech.}\ }%
  \textbf{\bibinfo {volume} {717}} (\bibinfo {year} {2013})%
  \bibAnnoteFile{NoStop}{vlad2013}%
\bibitem{riley}%
  \BibitemOpen
  \bibfield{author}{%
  \bibinfo {author} {\bibfnamefont{N.}~\bibnamefont{Riley}},\ }%
  \bibfield{journal}{%
  \bibinfo {journal} {Ann. Rev. Fluid Mech.}\ }%
  \textbf{\bibinfo {volume} {33}},\ \bibinfo {pages} {43} (\bibinfo {year}
  {2001})%
  \bibAnnoteFile{NoStop}{riley}%
\bibitem{SM}%
  \BibitemOpen
  \bibinfo {journal} {See Supplemental Materials for additional methods,
  movies, figures and references [25-28]}%
  \bibAnnoteFile{Stop}{SM}%
\bibitem{sorokin}
V. S. Sorokin, I. I. Blekhman and V. B. Vasilkov, Non-linear Dynamics {\bf 67}, 147 (2012).
\bibitem{mac}
F. H. Harlow and J. E. Welch, Phys. Fluids {\bf 8}, 2182 (1965).
\bibitem{projection}
K. H\"ofler and S. Schwarzer, Phys. Rev. E {\bf 61}, 7146 (2000). 
\bibitem{klotsa3}
K. D. Klotsa, PhD Thesis, University of Nottingham (2009).
\bibitem{otto}%
  \BibitemOpen
  \bibfield{author}{%
  \bibinfo {author} {\bibfnamefont{F.}~\bibnamefont{Otto}}, \bibinfo {author}
  {\bibfnamefont{E.~K.}\ \bibnamefont{Riegler}},\ and\ \bibinfo {author}
  {\bibfnamefont{G.~A.}\ \bibnamefont{Voth}},\ }%
  \bibfield{journal}{%
  \bibinfo {journal} {Phys. Fluids}\ }%
  \textbf{\bibinfo {volume} {20}},\ \bibinfo {pages} {093304} (\bibinfo {year}
  {2008})%
  \bibAnnoteFile{NoStop}{otto}%
\bibitem{klotsa1}%
  \BibitemOpen
\bibfield{journal}{%
    }%
  \bibfield{author}{%
  \bibinfo {author} {\bibfnamefont{D.}~\bibnamefont{Klotsa}}, \bibinfo {author}
  {\bibfnamefont{M.~R.}\ \bibnamefont{Swift}}, \bibinfo {author}
  {\bibfnamefont{R.~M.}\ \bibnamefont{Bowley}},\ and\ \bibinfo {author}
  {\bibfnamefont{P.~J.}\ \bibnamefont{King}},\ }%
  \bibfield{journal}{%
  \bibinfo {journal} {Phys. Rev. E}\ }%
  \textbf{\bibinfo {volume} {76}},\ \bibinfo {pages} {056314} (\bibinfo {year}
  {2007})%
  \bibAnnoteFile{NoStop}{klotsa1}%
\bibitem{wright}%
  \BibitemOpen
  \bibfield{author}{%
  \bibinfo {author} {\bibfnamefont{H.~S.}\ \bibnamefont{Wright}}, \bibinfo
  {author} {\bibfnamefont{M.~R.}\ \bibnamefont{Swift}},\ and\ \bibinfo {author}
  {\bibfnamefont{P.~J.}\ \bibnamefont{King}},\ }%
  \bibfield{journal}{%
  \bibinfo {journal} {Phys. Rev. E}\ }%
  \textbf{\bibinfo {volume} {78}},\ \bibinfo {pages} {036311} (\bibinfo {year}
  {2008})%
  \bibAnnoteFile{NoStop}{wright}%
\bibitem{klotsa2}%
  \BibitemOpen
  \bibfield{author}{%
  \bibinfo {author} {\bibfnamefont{D.}~\bibnamefont{Klotsa}}, \bibinfo {author}
  {\bibfnamefont{M.~R.}\ \bibnamefont{Swift}}, \bibinfo {author}
  {\bibfnamefont{R.~M.}\ \bibnamefont{Bowley}},\ and\ \bibinfo {author}
  {\bibfnamefont{P.~J.}\ \bibnamefont{King}},\ }%
  \bibfield{journal}{%
  \bibinfo {journal} {Phys. Rev. E}\ }%
  \textbf{\bibinfo {volume} {79}},\ \bibinfo {pages} {021302} (\bibinfo {year}
  {2009})%
  \bibAnnoteFile{NoStop}{klotsa2}%
\bibitem{klotsa3}%
  \BibitemOpen
  \bibfield{author}{%
  \bibinfo {author} {\bibfnamefont{K.~D.}\ \bibnamefont{Klotsa}},\ }%
  \bibfield{journal}{%
  \bibinfo {journal} {PhD Thesis}}%
   (\bibinfo {year} {2009})%
  \bibAnnoteFile{NoStop}{klotsa3}%
\bibitem{hector}%
  \BibitemOpen
  \bibfield{author}{%
  \bibinfo {author} {\bibfnamefont{H.~A.}\ \bibnamefont{Pacheco-Martinez}},
  \bibinfo {author} {\bibfnamefont{L.}~\bibnamefont{Liao}}, \bibinfo {author}
  {\bibfnamefont{R.~J.~A.}\ \bibnamefont{Hill}}, \bibinfo {author}
  {\bibfnamefont{M.~R.}\ \bibnamefont{Swift}},\ and\ \bibinfo {author}
  {\bibfnamefont{R.~M.}\ \bibnamefont{Bowley}},\ }%
  \bibfield{journal}{%
  \bibinfo {journal} {Phys. Rev. Lett.}\ }%
  \textbf{\bibinfo {volume} {110}},\ \bibinfo {pages} {154501} (\bibinfo {year}
  {2013})%
  \bibAnnoteFile{NoStop}{hector}%
\bibitem{mac}%
  \BibitemOpen
  \bibfield{author}{%
  \bibinfo {author} {\bibfnamefont{F.~H.}\ \bibnamefont{Harlow}}\ and\ \bibinfo
  {author} {\bibfnamefont{J.~E.}\ \bibnamefont{Welch}},\ }%
  \bibfield{journal}{%
  \bibinfo {journal} {Phys. Fluids}\ }%
  \textbf{\bibinfo {volume} {8}},\ \bibinfo {pages} {2182} (\bibinfo {year}
  {1965})%
  \bibAnnoteFile{NoStop}{mac}%
\bibitem{projection}%
  \BibitemOpen
  \bibfield{author}{%
  \bibinfo {author} {\bibfnamefont{K.}~\bibnamefont{H\"ofler}}\ and\ \bibinfo
  {author} {\bibfnamefont{S.}~\bibnamefont{Schwarzer}},\ }%
  \bibfield{journal}{%
  \bibinfo {journal} {Phys. Rev. E}\ }%
  \textbf{\bibinfo {volume} {61}},\ \bibinfo {pages} {7146} (\bibinfo {year}
  {2000})%
  \bibAnnoteFile{NoStop}{projection}%
\bibitem{belkin}%
  \BibitemOpen
  \bibfield{author}{%
  \bibinfo {author} {\bibfnamefont{M.}~\bibnamefont{Belkin}}, \bibinfo {author}
  {\bibfnamefont{A.}~\bibnamefont{Snezhko}}, \bibinfo {author}
  {\bibfnamefont{I.~S.}\ \bibnamefont{Aranson}},\ and\ \bibinfo {author}
  {\bibfnamefont{W.~K.}\ \bibnamefont{Kwok}},\ }%
  \bibfield{journal}{%
  \bibinfo {journal} {Phys. Rev. Lett.}\ }%
  \textbf{\bibinfo {volume} {99}},\ \bibinfo {pages} {158301} (\bibinfo {year}
  {2007})%
  \bibAnnoteFile{NoStop}{belkin}%
\bibitem{wun}%
  \BibitemOpen
  \bibfield{author}{%
  \bibinfo {author} {\bibfnamefont{R.}~\bibnamefont{Wunenburger}}, \bibinfo
  {author} {\bibfnamefont{V.}~\bibnamefont{Carrier}},\ and\ \bibinfo {author}
  {\bibfnamefont{Y.}~\bibnamefont{Garrabos}},\ }%
  \bibfield{journal}{%
  \bibinfo {journal} {Physics of Fluids}\ }%
  \textbf{\bibinfo {volume} {14}},\ \bibinfo {pages} {2350} (\bibinfo {year}
  {2002})%
  \bibAnnoteFile{NoStop}{wun}%
\bibitem{voth}%
  \BibitemOpen
  \bibfield{author}{%
  \bibinfo {author} {\bibfnamefont{G.~A.}\ \bibnamefont{Voth}}, \bibinfo
  {author} {\bibfnamefont{B.}~\bibnamefont{Bigger}}, \bibinfo {author}
  {\bibfnamefont{M.~R.}\ \bibnamefont{Buckley}}, \bibinfo {author}
  {\bibfnamefont{W.}~\bibnamefont{Losert}}, \bibinfo {author}
  {\bibfnamefont{M.~P.}\ \bibnamefont{Brenner}}, \bibinfo {author}
  {\bibfnamefont{H.~A.}\ \bibnamefont{Stone}},\ and\ \bibinfo {author}
  {\bibfnamefont{J.~P.}\ \bibnamefont{Gollub}},\ }%
  \bibfield{journal}{%
  \bibinfo {journal} {Phys. Rev. Lett.}\ }%
  \textbf{\bibinfo {volume} {88}},\ \bibinfo {pages} {234301} (\bibinfo {year}
  {2002})%
  \bibAnnoteFile{NoStop}{voth}%

\end{thebibliography}
%

\end{document}